\documentclass[twocolumn]{aa}
\usepackage{graphicx}
\usepackage{natbib}

\begin{document}
   \title{The X-ray soft excess in classical T Tauri stars}


   \author{M. G\"udel
          \inst{1,2}
          \and
          A. Telleschi
    	  \inst{1}
          }

   \offprints{M. G\"udel}

   \institute{Paul Scherrer Institut, W\"urenlingen and Villigen,
              CH-5232 Villigen PSI, Switzerland\\
              \email{guedel, atellesc@astro.phys.ethz.ch}
         \and
             Max-Planck-Institute for Astronomy, K\"onigstuhl 17, 
             69117 Heidelberg, Germany
             }

   \date{Received 22 June 2007; accepted 6 September 2007}

   \abstract
   {}
   {We study an anomaly in the X-ray flux (or luminosity) ratio between the O\,{\sc vii}\ $\lambda\lambda$21.6-22.1 
   triplet and the O\,{\sc viii} Ly$\alpha$ line seen in classical T Tauri stars (CTTS). 
   This ratio is unusually high when compared with
   ratios for main-sequence and non-accreting T Tauri stars  (Telleschi et al. 2007). 
   We compare these samples to identify the source of the excess. A sample of recently discovered X-ray stars with 
   a soft component attributed to jet emission is also considered.
    }
   {We  discuss data obtained from the {{\it XMM-Newton Extended Survey of the Taurus Molecular
   Cloud}} (XEST) project, complemented by data from the
   published literature. We also present data from the CTTS RU Lup.}
   {All CTTS in the sample show an anomalous O\,{\sc vii}/O\,{\sc viii} flux ratio when compared with
   WTTS or MS stars. The anomaly is due to an {{\it excess}} of cool, O\,{\sc vii} emitting material rather 
   than a deficiency of hotter plasma. The excess plasma must therefore have temperatures 
   of $\la 2$~MK.  This {{\it soft excess}} does not correlate with UV excesses of CTTS, but seems to be related
   with the stellar X-ray luminosity. The spectra of the jet-driving TTS do not fit into this picture.
   }
   {The soft excess  depends both on the presence of accretion streams in CTTS and on
   magnetic activity. The gas may be shock-heated near the surface, although it may also be
   heated in  the magnetospheric accretion funnels.  The soft component of the jet-driving sources is unlikely 
   to be due to the same process.
   }
   \keywords{Stars: coronae -- 
             Stars: formation --
             Stars: pre-main sequence --
             X-rays: stars }

   \maketitle
%

\section{Introduction}\label{intro}

Classical T Tauri stars (CTTS) are optically revealed young stars distinguished by spectra 
showing strong line emission of, e.g., H$\alpha$ or Ca\,{\sc ii} H\&K. 
Their H$\alpha$ and ultraviolet (UV) line and continuum emission is $10^2-10^4$ times stronger than  in
active main-sequence (MS) stars, regardless of 
the photospheric effective temperature but correlated with the mass accretion rates
derived from the optical continuum \citep{bouvier90, johnskrull00}. The  consensus, based on 
such correlations and line profiles, is that accreting material is heated in shocks near the 
stellar surface (e.g., \citealt{calvet98, gullbring98}). Some of the emission lines (e.g., H$\alpha$, 
Ca\,{\sc ii}) may also form in the accretion funnels or in stellar winds \citep{ardila02}.

Nearly free-falling gas will indeed heat  to maximum  
temperatures $T_s = 8.6\times 10^5$~K~$[M/(0.5M_{\odot})] [R/(2R_{\odot})]^{-1}$ in shocks \citep{calvet98}. 
UV and optical line emission thus diagnoses the accretion velocity, 
the mass accretion rate, and possibly the surface filling factor of accretion funnels. 
If the photoelectric absorption by the accreting gas is small, then the softest  
X-ray range may reveal the high-$T$ tail of the shock emission measure. 
\citet{guedel06}, \citet{telleschi07c} and \citet{guedel07c} were first to identify 
an excess in the O\,{\sc vii}/O\,{\sc viii} Ly$\alpha$ flux (or luminosity) ratio in CTTS when 
compared with MS stars, or weak-line T Tauri stars (WTTS) in the {\it XEST} project  \citep{guedel07a}, 
the so-called X-ray {\it soft excess} of CTTS.  This feature may define the high-$T$ 
continuation of the excess emission diagnosed for the $10^4-10^5$~K range. 

The appreciable accretion rates predict shock densities of $n_{\rm e}\approx 10^{12}- 10^{14}$~cm$^{-3}$, as 
first indeed reported for the CTTS TW Hya  from line diagnostics of  O\,{\sc vii} and Ne\,{\sc ix},
forming at a few MK \citep{kastner02, stelzer04}.  However, some accreting
young stars show much lower $n_{\rm e}$, such as AB Aur \citep{telleschi07b} and 
T Tau \citep{guedel07c}; the same discrepancy between expected and observed $n_{\rm e}$ has also 
been reported from UV density diagnostics \citep{johnskrull00}.

Here, we present new evidence for the soft excess and discuss it in a larger 
context. We are interested in comparing CTTS with WTTS
to assess whether the anomaly can be attributed to accretion in CTTS; we will therefore include WTTS
as a test sample and compare with samples of MS stars. Further, we will test whether the anomalous
O\,{\sc vii}/O\,{\sc viii} flux ratio is due to a high O\,{\sc vii} flux or rather due to a suppressed
O\,{\sc viii}~Ly$\alpha$ flux. We compare these stellar samples with a small group of accreting,
jet-driving stars that also show an anomalous soft component \citep{guedel07b}. We also compare
the X-ray soft excess with UV line and continuum excesses reported earlier.

\section{Data and Results}

Ratios between fluxes in
the O\,{\sc vii} triplet ($\lambda\lambda 21.6, 21.8, 22.1$, summed over all three lines) 
and the O\,{\sc viii}  Ly$\alpha$ $\lambda$18.97 line have been derived from fluxes from
\citet{telleschi07c} (for the WTTS HD~283572, V773 Tau, V410 Tau, and HP Tau/G2; for the CTTS 
SU Aur, DH Tau, DN Tau, BP Tau; and for the Herbig star AB Aur), \citet{guedel07c} (for the CTTS 
T Tau N), \citet{robrade06} (for the CTTS BP Tau, TW Hya, and CR Cha), \citet{argiroffi05} 
(for the multiple CTTS-WTTS system TWA~5; it is unknown whether the CTTS or the WTTS is the
dominant X-ray source), \citet{guenther06} (for the CTTS V4046 Sgr), and
\citet{argiroffi07} (for the old CTTS MP Mus), and our own analysis for the CTTS
RU Lup.\footnote{{\it XMM-Newton} ObsID~0303900301, observed on 2005 August 8 during 29.8~ks; 
data reduction followed the same procedures as described by \citet{guedel07a} and  \citet{telleschi07c};
the line fluxes were found from spectral fits based on the apec model in the XSPEC software package 
\citep{arnaud96}. Key results:
Three spectral components with $kT_1 = 0.24$~keV, $kT_2 = 0.92$~keV, and $kT_3 = 3.6$~keV; 
emission measure ratio = 1.0:2.9:1.5; $N_{\rm H} = 1.6\times 10^{21}$~cm$^{-2}$.
Luminosity of the O\,{\sc vii} triplet: $(4.0\pm 1.4)\times 10^{28}$~erg~s$^{-1}$; 
of the O\,{\sc viii}~Ly$\alpha$ line: $(4.1\pm 1.4)\times 10^{28}$~erg~s$^{-1}$;
total X-ray [0.3~keV, 10~keV] luminosity: $L_{\rm X} = 1.9\times 10^{30}$~erg~s$^{-1}$. The distance
to RU Lup is 140~pc \citep{bertout99}.}

We determined the unabsorbed fluxes by correcting for wavelength-dependent transmission,
as calculated in XSPEC 
using the ``wabs'' model. We used the absorption column densities,  $N_{\rm H}$, toward the target
stars, also given by the above authors. We adopted $N_{\rm H} = 10^{21}$~cm$^{-3}$ for V4046 Sgr, as 
suggested by \citet{guenther06}. Finally, line luminosities were calculated using the published distances.

Total X-ray luminosities, $L_{\rm X}$, were taken
from the same authors; for V4046 Sgr, the spectral line fluxes are about half
as high as for TW Hya (see comparison in \citealt{guenther06}) while the distance seems to 
be very uncertain; we adopted a distance of 83~pc \citep{quast00};  the uncertainties
will not be crucial for our investigation. For MP Mus, distance and $L_{\rm X}$ are from 
\citet{mamajek02}.

Data for solar-analog (G-type) MS stars were taken from \citet{telleschi05}, and for a larger
MS sample from \citet{ness04}. These authors list $L_{\rm X}$, 
the energy fluxes (in erg~cm$^{-2}$~s$^{-1}$)
in the O\,{\sc viii} Ly$\alpha$ line and in the O\,{\sc vii} He$\alpha$ triplet (only the $r$ line
in \citealt{ness04}), from which $L$(O\,{\sc viii}) resp. $L$(O\,{\sc vii})   were calculated.
For these stars, $N_{\rm H}$ is low and does not need to be considered.

\begin{figure}[t!]
\hbox{
\includegraphics[angle=-0,width=8.8cm]{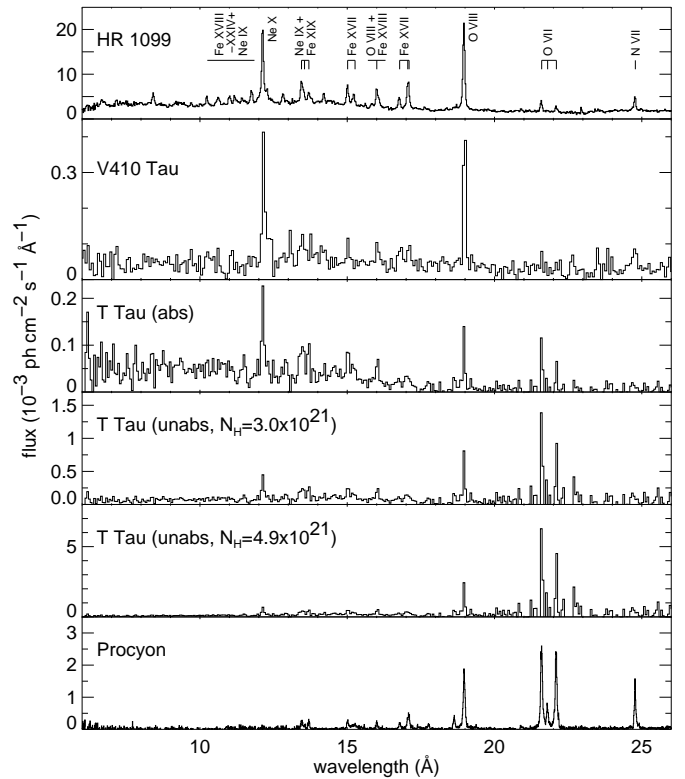}
}
\vskip -0.3truecm
\caption{Comparison of fluxed X-ray photon spectra of (from top to bottom, all from {\it XMM-Newton RGS}) the active binary HR~1099, 
         the WTTS V410~Tau, the CTTS T Tau, the T Tau spectrum modeled after removal of absorption (two versions,  
	 using $N_{\rm H} = 3\times 10^{21}$~cm$^{-2}$ and $4.9\times 10^{21}$~cm$^{-2}$, respectively), 
	 and the inactive MS star Procyon.  
	 The bins are equidistant in wavelength (from top to bottom, the bin widths are, respectively, 
	 0.025~\AA, 0.070~\AA, 0.058~\AA, 0.058~\AA, 0.058~\AA, and 0.010~\AA). 
	 }
\label{fig1} 
\end{figure}

Fig.~\ref{fig1} highlights the soft excess for the CTTS T Tau N. The figure compares the X-ray spectrum
of the active, evolved binary HR~1099 dominated by emission from a K-type subgiant (top panel; archival 
{\it XMM-Newton} data, see \citealt{audard01}) 
with the spectrum of the weakly absorbed WTTS V410~Tau \citep{telleschi07c}, the CTTS T Tau 
\citep{guedel07c}, and the old single F-type star Procyon (archival  {\it XMM-Newton} data, see 
\citealt{raassen02}). HR~1099 and V410~Tau show the typical signatures of a hot, active
corona: a strong continuum, strong lines of Ne\,{\sc x}, and highly-ionized Fe lines but little flux
in the O\,{\sc vii} line triplet. In contrast, the spectrum of Procyon is dominated by lines of C, N, and O, the 
O\,{\sc vii} triplet exceeding the O\,{\sc viii} Ly$\alpha$ line in flux. The observed spectrum of T Tau reveals
a hybrid situation, with signatures of a very active corona shortward of 19~\AA\ but also an unusually strong
O\,{\sc vii} triplet. Because its hydrogen absorption is large (in contrast to V410 Tau - note the latter's 
N\,{\sc vii} Ly$\alpha$ $\lambda$24.8 line formed over a wide temperature range), 
we have modeled the intrinsic, unabsorbed
spectrum based on transmissions determined in XSPEC (based on the ``wabs'' model) using $N_{\rm H}$ from 
\citet{guedel07c} ($N_{\rm H} = 4.9\times 10^{21}$~cm$^{-2}$), but also the somewhat lower value found 
from EPIC spectra ($N_{\rm H} \approx 3\times 10^{21}$~cm$^{-2}$; \citealt{guedel07a}). In either case,
the O\,{\sc vii} lines are now the strongest lines in the X-ray spectrum, reminiscent of the situation in  Procyon.

\begin{figure}[t!]
\hbox{
\includegraphics[angle=-0,width=8.8cm,height=7.2cm]{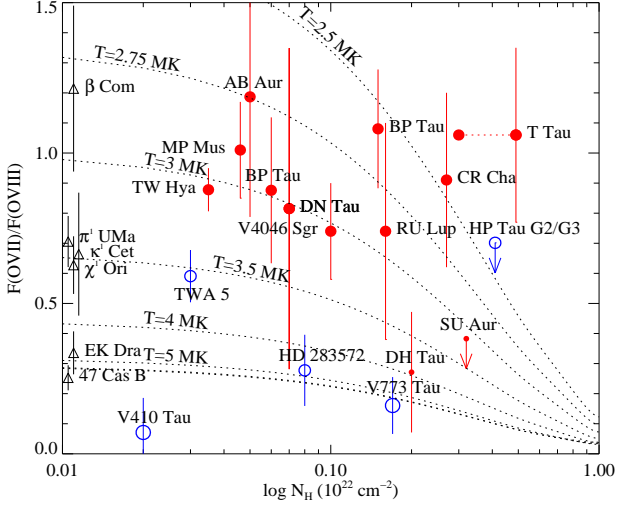}
}\vskip -0.1truecm
\caption{The ratio between O\,{\sc vii} and O\,{\sc viii} Ly$\alpha$ photon fluxes (each in units of ph~cm$^{-2}$~s$^{-1}$, 
         absorbed)  for CTTS (red, filled circles) and WTTS (blue, open circles)
         versus  $N_{\rm H}$. Two flaring CTTS are shown by small, filled circles. 
	 The loci of MS solar analogs are shown by black triangles near $N_{\rm H} \approx 0.01$ 
	 although their true $N_{\rm H}$ are much lower. The dotted lines give the 
	 ratios for isothermal plasmas with temperatures as labeled. BP Tau is plotted twice, 
	 once after \citet{robrade06} (larger $N_{\rm H}$)  and once after \citet{telleschi07c}.
	 The high- and low-absorption solutions for T Tau are connected by a dotted line.
	 }
\label{fig2} 
\end{figure}

Fig.~\ref{fig2}  shows the measured (absorbed)
O\,{\sc vii}/O\,{\sc viii} Ly$\alpha$ photon-flux ratio, $S$, versus the X-ray determined $N_{\rm H}.$\footnote{DN Tau 
has been reconsidered, with very similar results as in \citet{telleschi07c}; for SU Aur and HP Tau G2/G3, upper 
limits to the O\,{\sc vii} flux derived by \citet{telleschi07c} [see also their Fig.~8] have been adopted; 
BP Tau is plotted twice, given the discrepant  $N_{\rm H}$ values between \citet{robrade06} and \citet{telleschi07c}.} 
CTTS and WTTS are marked, respectively, by filled (red) and open (blue) circles;
the flaring CTTS SU Aur and DH Tau  are marked by small, filled circles. The dotted lines mark
the loci of $S$ for an isothermal plasma; the labels give the
electron temperatures. The plasma contributing to O\,{\sc vii} and O\,{\sc viii} is, however,  not 
isothermal as the hotter plasma also contributes to the O\,{\sc viii} flux. All CTTS show 
$S \approx 1 \pm  0.25$ in this range of $N_{\rm H}$. WTTS are found at much lower values; their
published spectra show at best marginal evidence of the O\,{\sc vii} triplet 
(e.g., \citealt{telleschi07c}). 

MS solar analogs  are plotted near $N_{\rm H} = 0.01$ although their true $N_{\rm H}$ are much lower. 
Characteristic coronal temperatures, $T$, of MS stars are a function of $L_{\rm X}$. For 
solar analogs at different activity levels, $T$ increases from $\approx 2$~MK 
for a solar twin with $L_{\rm X} \approx 10^{27}$~erg~s$^{-1}$  to $T \approx 10$~MK for a ZAMS star with 
$L_{\rm X} = (1-3)\times 10^{30}$~erg~s$^{-1}$ \citep{telleschi05}. Consequently, $S$ is expected to decrease
toward higher activity levels (i.e., the  O\,{\sc vii} line becomes progressively less important; Fig.~\ref{fig1}). 
This is indeed the case, the most active solar analog in the sample, 47 Cas B, showing $S \approx 0.26$, and
the least active one, $\beta$~Com, $S\approx 1.21$.
The values of $S$ for WTTS and the two active, near-ZAMS stars 47 Cas B and EK Dra are similar, whereas 
for CTTS they are similar to inactive solar analogs with ages of $\approx 1$~Gyr.

\begin{figure}[t!]
\vbox{
\includegraphics[angle=-0,width=8.cm]{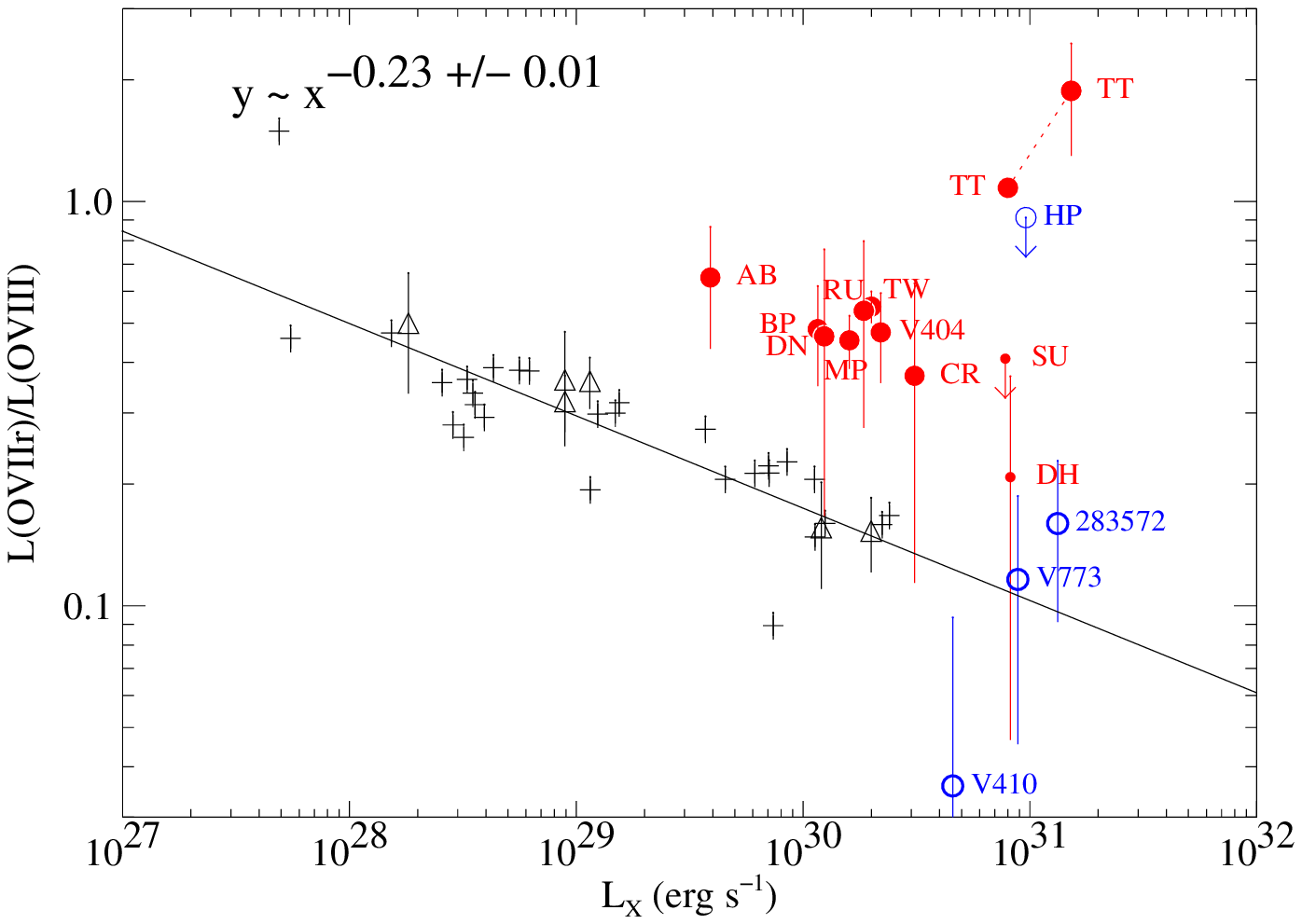}
\includegraphics[angle=-0,width=8.cm]{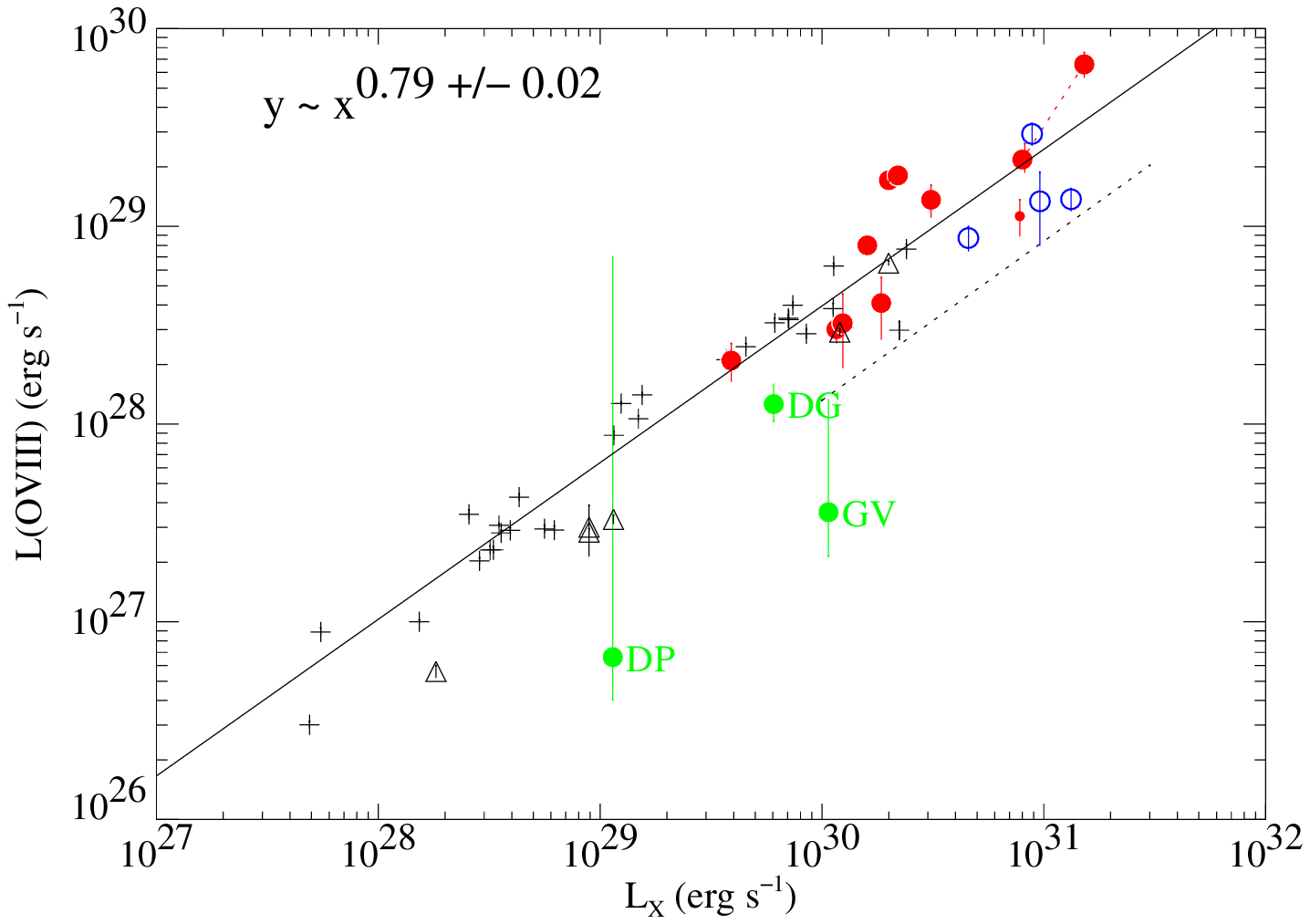}
\includegraphics[angle=-0,width=8.cm]{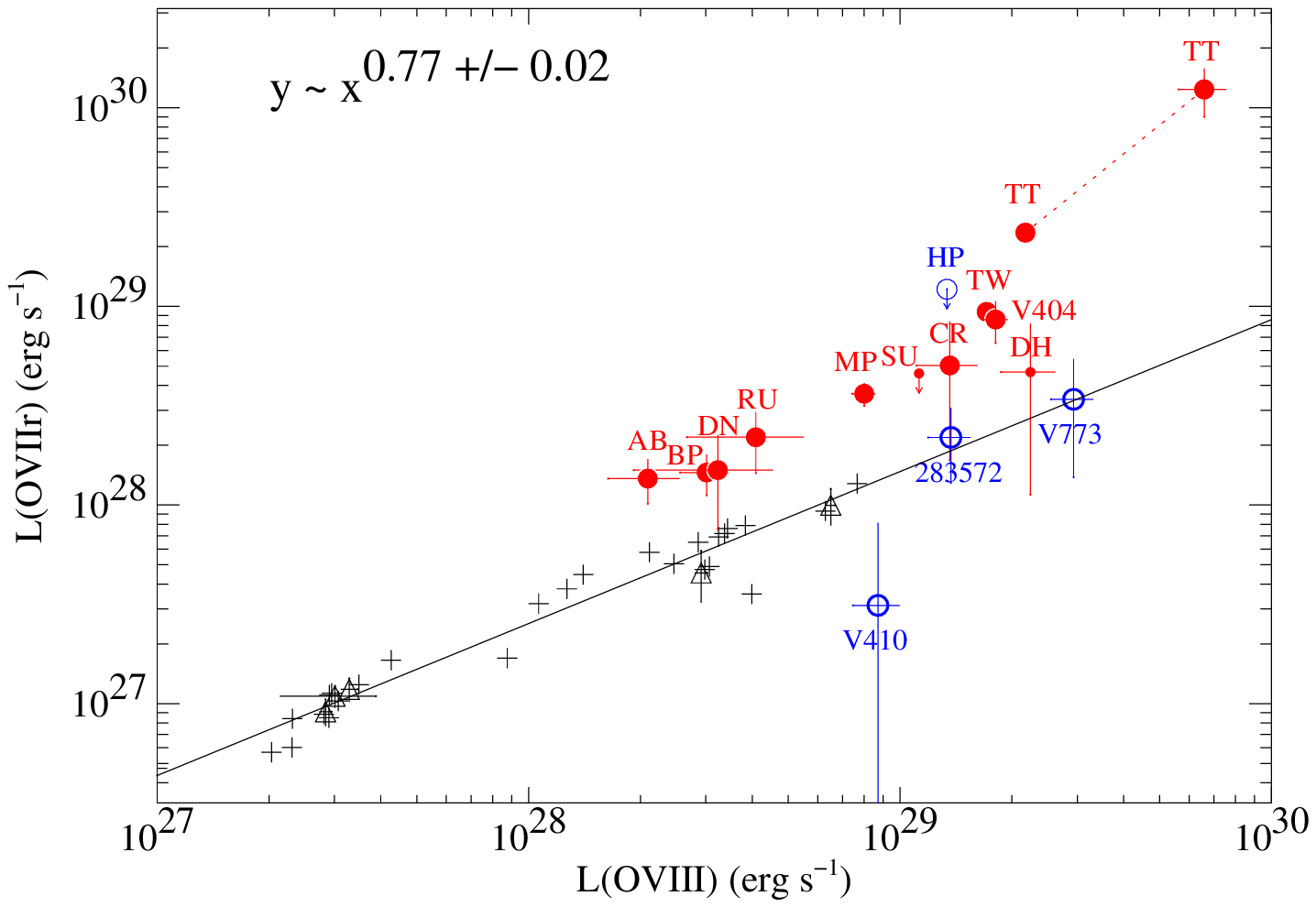}
}
\caption{{\it Top (a):} The ratio between O\,{\sc vii} $r$ and O\,{\sc viii} Ly$\alpha$ luminosities (each in 
units of erg~s$^{-1}$) vs. $L_{\rm X}$.  Labels give initial letters of stellar names. Crosses mark MS stars 
(from \citealt{ness04}), triangles solar analogs (from \citealt{telleschi05}), 
filled (red) circles CTTS, and open (blue) circles WTTS. The high- and low-absorption solutions 
for T Tau are connected by a dotted line. The solid line is a power-law fit to the 
MS stars with $L_{\rm X} > 10^{27}$~erg~s$^{-1}$ (the  relation between the variables is given in the upper left 
corner). -- {\it Middle (b):} Correlation between $L$(O\,{\sc viii})  and $L_{\rm X}$ for MS stars. Symbols are as in 
Fig.~\ref{fig3}a. All three samples follow the same trend (solid line for MS stars). The dotted line corresponds to a 
suppression of $L$(O\,{\sc viii}) by a factor of 3. Also shown are three TAX sources (filled, green, 
labeled: DG Tau, GV Tau, DP Tau). --
{\it Bottom (c):} Correlation between $L$(O\,{\sc vii} $r$) and $L$(O\,{\sc viii}). Symbols are as in Fig.~\ref{fig3}a.   
}
\label{fig3} 
\end{figure}

Fig.~\ref{fig3}a shows the ratio between the intrinsic (unabsorbed) luminosities of the O\,{\sc vii} $r$ 
and the O\,{\sc viii} Ly$\alpha$ lines as a function of $L_{\rm X}$, comparing with solar analogs from 
\citet{telleschi05} and  the larger MS sample from \citet{ness04}. 
For the TTS of \citet{telleschi07c}, a good approximation to compute $L$(O\,{\sc vii} $r$) 
from $L$(O\,{\sc vii}) is $L$(O\,{\sc vii} $r$) $= 0.55L$(O\,{\sc vii}) \citep{porquet01}.\footnote{We cannot 
strictly use the $S$ ratio to derive the fractional flux of the O\,{\sc vii} $r$ line 
\citep{porquet01} because the plasma is unlikely to be isothermal.}
The trend of a decreasing ratio with increasing $L_{\rm X}$ for MS stars is followed by the sample of WTTS, while 
CTTS again show a significant excess. This also holds if the the surface X-ray flux 
($L_{\rm X}/[4\pi R_*^2]$) is used on the abscissa (not shown).

In principle, a suppressed O\,{\sc viii} Ly$\alpha$ flux
in CTTS would produce the same anomaly, but as we show in Fig.~\ref{fig3}b, $L$(O\,{\sc viii}) 
of CTTS and WTTS both follow the same trend (rather than being suppressed for CTTS by a factor of $\approx$3--4), 
also indistinguishable from MS stars. Therefore, the CTTS
{\it soft excess} is due to an excess in the O\,{\sc vii} flux.

We also consider a sample of strongly accreting {\it Two-Absorber X-ray} (TAX) sources that show an 
anomalous soft component additional to a much more strongly absorbed coronal component; the soft component 
contributes essentially all of the observed O\,{\sc vii} and O\,{\sc viii}~Ly$\alpha$ flux; it has been 
interpreted as originating from the base of jets \citep{guedel07b}. The low-resolution EPIC spectra 
are not useful to derive an $S$ ratio, but the spectral fits are appropriate to estimate the luminosity in
the dominant O\,{\sc viii}~Ly$\alpha$ line from the model. Overplotting $L$(O\,{\sc viii})  
in Fig.~\ref{fig3}b shows a poor correlation, as should be expected because the coronal component shows a
separate (harder) spectrum whose  O\,{\sc viii}~Ly$\alpha$ line is, however, entirely absorbed.

\section{Discussion and conclusions}

All CTTS (except the two flaring ones) show an {\it X-ray soft excess} defined
by an anomalously high ratio between the fluxes of the O\,{\sc vii} triplet and the
O\,{\sc viii}~Ly$\alpha$  line, compared to WTTS and MS stars.  The anomaly refers to CTTS only, while WTTS show 
line ratios comparable with very active MS stars.
In contrast, the correlation between $L$(O\,{\sc viii}) and $L_{\rm X}$ is indistinguishable  
between CTTS, WTTS, and MS stars, suggesting that the CTTS soft excess is indeed due to an excess 
of cool material with $T \la 2$~MK rather than a relative suppression of the  
O\,{\sc viii}~Ly$\alpha$ line (e.g., due to resonance scattering). 
The strongly accreting TAX sources  do not fit into this picture; strong absorption makes
their {\it coronal} O\,{\sc vii} and O\,{\sc viii}~Ly$\alpha$ lines
inaccessible. 

Is the X-ray {\it soft excess} described here the high-temperature equivalent of the UV/optical 
continuum and  line excesses? We correlated $L$(O\,{\sc vii} $r$), which is dominated by the excess emission, with  
UV Si\,{\sc ii}, Si\,{\sc iv}, C\,{\sc iv}, and $\lambda$1958 continuum luminosities (all strongly 
dominated by excess emission) derived from fluxes presented by \citet{valenti00}, applying the 
$R=3.1$ extinction law given by \citet{cardelli89}.
No correlation was found. This is little surprising given that the UV excess luminosities amount to
$10^2-10^4$ times the levels of ``normal'' stars, with a large scatter over the entire range 
\citep{johnskrull00}, while the  O\,{\sc vii} excess is comparatively small. Fig.~\ref{fig3} in fact
suggests that for most CTTS reported here, the $L$(O\,{\sc vii} $r$)/$L$(O\,{\sc viii}) ratio is a 
factor of $\approx$3--4 that of equivalent
MS stars, i.e., $L$(O\,{\sc vii} $r$) scales with  $L$(O\,{\sc viii})  as 
in MS stars, except that it is enhanced by a factor of $\approx$3--4. $L$(O\,{\sc viii})  
in turn is ``normal'' and scales with $L_{\rm X}$ (Fig.~\ref{fig3}b). Because O\,{\sc vii} and O\,{\sc viii} have overlapping
formation temperature ranges, their flux ratio cannot take arbitrary values for reasonable EM distributions. 
However, much larger (but also smaller) excesses are easily possible given the variations in the UV excess.
For very cool excess plasma, the added O\,{\sc vii} flux could be small and the correlation with the UV excess
be masked by scatter, but then the O\,{\sc vii} would be close to the ``normal'' coronal level.

The $L$(O\,{\sc vii} $r$) vs. $L$(O\,{\sc viii}) relation
is shown in Fig.~\ref{fig3}c (to avoid larger error bars, we have used $L$(O\,{\sc vii} $r$) without 
the minor correction for the non-excess contribution). This trend does not connect to the behavior of the UV 
excess at {\it lower} $T$ but is reminiscent of the systematic factor $\approx 2$ deficiency of
(harder) X-rays from {\it hotter} plasma in CTTS with respect to WTTS \citep{telleschi07a}.
It appears that the {\it X-ray soft excess depends on the level of magnetic (``coronal'') activity.}
At the same time,  {\it it depends on the presence of accretion}. 

Magnetic accretion streams may shock-heat gas at the impact
point in the stellar photosphere to X-ray emitting temperatures. If  $L$(O\,{\sc vii} $r$) $\propto
\dot{M}$ (similarly to the UV excess, \citealt{johnskrull00}), and since $\dot{M} \propto M^2$ \citep{muzerolle03}, 
$L_{\rm X} \propto M^{1.7\pm 0.2}$ \citep{telleschi07a}, and $L$(O\,{\sc viii}) $\propto L_{\rm X}^{0.79\pm 0.02}$ 
(Fig.~\ref{fig3}b), one might expect $L$(O\,{\sc vii} $r$) $\propto L$(O\,{\sc viii})$^{1.49\pm 0.18}$
although this is not a physical relationship. This relation is steeper than in Fig.~\ref{fig3}c; also, a direct 
correlation with the UV excess (which itself correlates with $\dot{M}$; \citealt{johnskrull00}) 
is not seen.  

On the other hand, the two dependencies may point to an interaction between 
accretion and magnetic activity. The cool, infalling material
may partly cool pre-existing heated coronal plasma, or reduce the efficiency of coronal heating in 
the regions of infall \citep{preibisch05, telleschi07c, guedel07c}. A model of this kind in which the soft 
excess is formed {\it in the accretion funnels} by the coronal heating process would make a 
correlation with the (coronal) $L_{\rm X}$  plausible, and would at the same time explain why CTTS 
are statistically X-ray fainter compared with WTTS \citep{preibisch05, 
telleschi07a}. The deficit of hot material may then reflect in additional, cooled
(or less heated) plasma evident in the soft excess in CTTS, i.e., the soft excess connects to the
hotter, coronal EM, as observed,  rather than to the cooler UV excess. Either way, it
clearly argues in favor of a substantial  influence of accretion on the X-ray properties of 
pre-main sequence stars.

\begin{acknowledgements}
We thank our referee, Fred Walter, for stimulating suggestions that have improved this 
paper. Kevin Briggs is acknowledged for help in the data reduction. The International Space Science Institute 
(ISSI) in Bern financially supported  the {\it XEST} project. This research is based on observations obtained 
with {\it XMM-Newton}, an ESA science mission
with instruments and contributions directly funded by ESA member states and the USA (NASA).  
\end{acknowledgements}


\begin{thebibliography}{}
\bibitem[Ardila et al.(2002)]{ardila02}Ardila, D., Basri, G., Walter, F.~M., Valenti, J.~A., \&
         Johns-Krull, C.~M. 2002, ApJ, 567, 1013
\bibitem[Argiroffi et al.(2005)]{argiroffi05}Argiroffi, C., Maggio, A., \& Peres, G., et al. 2005, 
         A\&A, 439, 1149
\bibitem[Argiroffi et al.(2007)]{argiroffi07}Argiroffi, C., Maggio, A., \& Peres, G. 2007, A\&A, 465, L5
\bibitem[Arnaud(1996)]{arnaud96}Arnaud, K.~A. 1996, in Astronomical Data Analysis Software and
         Systems V, ed. G. Jacoby \& J. Barnes (San Francisco: ASP),  17
\bibitem[Audard et al.(2001)]{audard01}Audard, M., G\"udel, M., \& Mewe, R. 2001, A\&A, 365, L318
\bibitem[Bertout et al.(1999)]{bertout99}Bertout, C., Robichon, N., \& Arenou, F. 1999, A\&A, 352, 574
\bibitem[Bouvier(1990)]{bouvier90}Bouvier, J. 1990, AJ, 99, 946
\bibitem[Calvet \& Gullbring(1998)]{calvet98}Calvet, N., \& Gullbring, E. 1998, ApJ, 509, 802
\bibitem[Cardelli et al.(1989)]{cardelli89}Cardelli, J.~A., Clayton, G.~C., \& Mathis, J.~S. 1989, 
         ApJ, 345, 245
\bibitem[G\"udel(2006)]{guedel06}G\"udel, M. 2006, in High-Resolution X-Ray Spect\-ros\-copy;
         http:// www.mssl.ucl.ac.uk/$\sim$gbr/workshop2/, astro-ph/0609281
\bibitem[G\"udel et al.(2007a)]{guedel07a}G\"udel, M., Briggs, K.~R., Arzner, K., et al. 2007a, A\&A, 468, 353  
\bibitem[G\"udel et al.(2007b)]{guedel07b}G\"udel, M., Telleschi, A., Audard, M.,  et al. 2007b, A\&A, 468, 515  
\bibitem[G\"udel et al.(2007c)]{guedel07c}G\"udel, M., Skinner, S.L., Mel'nikov, S.Y., et al. 2007c, A\&A, 468, 529  
\bibitem[Gullbring et al.(1998)]{gullbring98}Gullbring, E., Hartmann, L., Brice\~no, C., \& Calvet, N.
         1998, ApJ, 492, 323
\bibitem[G\"unther et al.(2006)]{guenther06}G\"unther, H.M., Liefke, C., \& Schmitt, J.H.M.M.  
          2006, A\&A, 459, L29
\bibitem[Johns-Krull et al.(2000)]{johnskrull00}Johns-Krull, C.M., Valenti, J.A., \& Linsky, J.L.
         2000, ApJ, 539, 815
\bibitem[Kastner et al.(2002)]{kastner02}Kastner, J.~H., Huenemoerder, D.~P., Schulz, N.~S., Canizares, C.~R.,
         \& Weintraub, D.~A. 2002, ApJ, 567, 434
\bibitem[Mamajek et al.(2002)]{mamajek02}Mamajek, E.~E., Meyer, M.~R., \& Liebert, J. 2002, AJ, 124, 1670
\bibitem[Muzerolle et al.(2003)]{muzerolle03}Muzerolle, J., Hillenbrand, L., Calvet, N., Brice\~no, C., \& Hartmann, L. 2003, ApJ, 
         592, 266
\bibitem[Ness et al.(2004)]{ness04}Ness, J.-U., G\"udel, M., Schmitt, J.~H.~M.~M., Audard, M., 
         \& Telleschi, A. 2004, A\&A, 427, 667
\bibitem[Preibisch et al.(2005)]{preibisch05}Preibisch, T., Kim, Y.-C., Favata, F., et al. 2005, ApJS, 160, 401 
\bibitem[Porquet et al.(2001)]{porquet01}Porquet, D., Mewe, R. Dubau, J., Raassen, A.~J.~J., \& Kaastra, J.~S. 2001, A\&A, 376, 1113
\bibitem[Quast et al.(2000)]{quast00}Quast, G.~R.,  Torres, C.~A.~O., de La Reza, R., da Silva, L.,
         \& Mayor, M. 2000, IAU Symp. 200, eds. B. Reipurth \&  H. Zinnecker, 28
\bibitem[Raassen et al.(2002)]{raassen02}Raassen, A.~J.~J., Mewe, R., Audard, M., et al. 2002, A\&A, 389, 228
\bibitem[Robrade \& Schmitt(2006)]{robrade06}Robrade, J., \& Schmitt, J.H.M.M. 2006, A\&A, 449, 737
\bibitem[Stelzer \& Schmitt(2004)]{stelzer04}Stelzer, B., \& Schmitt, J.~H.~M.~M. 2004, A\&A, 418, 687 
\bibitem[Telleschi et al.(2005)]{telleschi05}Telleschi, A., G\"udel, M., Briggs, K.~R., et al. 2005, ApJ, 622, 653  
\bibitem[Telleschi et al.(2007a)]{telleschi07a}Telleschi, A., G\"udel, M., Briggs, K.~R., et al. 2007a, A\&A, 468, 425  
\bibitem[Telleschi et al.(2007b)]{telleschi07b}Telleschi, A., G\"udel, M., Briggs, K.~R., et al. 2007b, A\&A, 468, 541  
\bibitem[Telleschi et al.(2007c)]{telleschi07c}Telleschi, A., G\"udel, M., Briggs, K.~R., et al. 2007c, A\&A, 468, 443  
\bibitem[Valenti et al.(2000)]{valenti00}Valenti, J.A., Johns-Krull, C.M., \& Linsky, J.L.
         2000, ApJS, 129, 399
\end{thebibliography}
\end{document}